\documentclass{sig-alternate}
\usepackage{amsmath}
\usepackage{verbatim}
\usepackage{amssymb}
\usepackage{graphicx}
\usepackage{mdwlist}
\usepackage{cite}
\usepackage{tabularx}
\usepackage{hyperref}
\usepackage{balance}
\usepackage{booktabs}
\usepackage{dsfont}
\usepackage{multirow}
\usepackage[T1]{fontenc}
\usepackage[subrefformat=parens,labelformat=parens,caption=false]{subfig}
\usepackage{pifont}

\newcommand{\eat}[1]{}

\newcommand{\lyxdot}{.}

\newcommand{\cmark}{\ding{51}}
\newcommand{\xmark}{-}

\renewcommand{\P}{\mathbb{P}}

\begin{document}
\makeatletter
\def\@copyrightspace{\relax}
\makeatother
%
\title{Modeling Website Popularity Competition in the \\ Attention-Activity Marketplace}

\numberofauthors{2} 

\author{
%
%
\alignauthor
Bruno Ribeiro\\
\affaddr{Carnegie Mellon University}\\
\affaddr{School of Computer Science}\\
\affaddr{Pittsburgh, PA, USA.}\\
    \email{ribeiro@cs.cmu.edu}
\and
    \alignauthor
    Christos Faloutsos\\
\affaddr{Carnegie Mellon University}\\
\affaddr{School of Computer Science}\\
\affaddr{Pittsburgh, PA, USA.}\\
    \email{christos@cs.cmu.edu}
}

\maketitle
\begin{abstract}
How does a new startup drive the popularity of competing websites into oblivion like Facebook famously did to MySpace? This question is of great interest to academics, technologists,  and financial investors alike. In this work we exploit the singular way in which Facebook wiped out the popularity of MySpace, Hi5, Friendster, and Multiply to guide the design of a new popularity competition model. Our model provides new insights into what Nobel Laureate Herbert A.\ Simon called the ``marketplace of attention,'' which we recast as the {\em attention-activity marketplace}. Our model design is further substantiated by user-level activity of 250,000 MySpace users obtained between 2004 and 2009. The resulting model not only accurately fits the observed Daily Active Users (DAU) of Facebook and its competitors but also predicts their fate four years into the future. 
\end{abstract}


\section{Introduction}


\newcommand{\ra}[1]{\renewcommand{\arraystretch}{#1}}

Membership-based websites such as Facebook are a proven success in what the late Nobel Laureate Herbert A.\ Simon called ``the marketplace of attention.''
In a 1969 lecture~\cite{Simon} Simon observed that many information systems were designed as if information was scarce, when the problem is just the opposite: ``[...] in an information-rich world, the wealth of information means a dearth of something else: a scarcity of whatever it is that information consumes. What information consumes is rather obvious: it consumes the attention of its recipients.''
In this context, Facebook, its competitors, and other analogous membership-based websites can be understood as Simon's {\em information processing systems} that speak more than they help us listen (digest information)~\cite{Simon}.

Casting the website popularity competition (e.g.\ Facebook v.s.\ MySpace) into Simon's insightful framework, we note that membership-based websites have an extra element: user attention is converted into content through user activity, which in turn consumes the attention of other users, thus creating an  {\em attention-activity marketplace}.
This observation inspires the following set of questions:
\vspace{-3pt}
\begin{itemize}
\item Can the attention-activity marketplace help explain the death of MySpace, Hi5, Friendster and Multiply\footnote{Section~\ref{s:datasets} provides a brief history of the Facebook, MySpace, Hi5, Friendster and Multiply websites.}?
\item Was Facebook the likely reason why MySpace, Hi5, Friendster, and Multiply popularity dwindled or did they die of ``natural causes'', e.g., ``users were bored''?
\item More broadly, is it possible to model the dynamics of user attention and activity as to capture the phenomenon that drives down the popularity of well established websites (e.g.\ MySpace)?  
\item Is it possible to learn the parameters of such model using widely available Daily Active Users (DAU) time series?
\end{itemize}

\subsection*{Model}
In this work we take a positive step toward answering the above questions.
We use the puzzling way by which Facebook has interfered with the popularity of its competitors on July 2008 to put forth a set of desired model properties.
One of the key desired properties is modeling user attention as a scarce resource that must be consumed by websites or other online user activities.
With the captured share of attention a website engages a user in content creation (activity), which in turn captures the attention of other users.
Users also have other online interests aside from the website and these interests also compete for attention (a competing website or other online activities).
The popularity competition ensues when two websites fight for the attention of their concurrent users (the {\em concurrent adopters} of both websites).
Media, marketing, and word-of-mouth adoptions complement the model as the driving forces behind membership growth.

The resulting model is a compartmental reaction-diffusion population-level model that provides DAU popularity forecasts and
offers a compelling hypothesis for the popularity competition of membership-based websites such as Facebook, MySpace, Hi5, Friendster, and Multiply.
Our model is an attention-activity market generalization of our previous single website model in Ribeiro~\cite{Ribeiro:2014www}.
The resulting  code of the algorithm is available online\footnote{\small\url{http://www.cs.cmu.edu/~ribeiro/Ribeiro_F_AttComp.zip}}.
\subsection*{Agreement with Observed Data}
Our modeling framework reveals interesting traits in our data. 
Websites almost universally have low barriers of adoption, as opposed to adoptions in consumer products such as smartphones and laptop computers that have a steep barrier to adopting multiple products. Signing up for Facebook and the like is free and reasonably effortless.
The main resource consumed by these websites is our time. Not coincidentally {\em time} is H.A.\ Simon's choice of attention metric~\cite{Simon}. 
As long as we see value in spending our time at these websites -- rather than somewhere else -- Facebook, MySpace, Hi5, Friendster, and Multiply can co-exist without interfering with each other.

\begin{figure}[!!bht]
\centering
\subfloat[][Facebook DAU time series showing insignificant DAU bump near July 20, 2008.\label{f:fb}]{
\includegraphics[width=2.6in,height=2.0in]{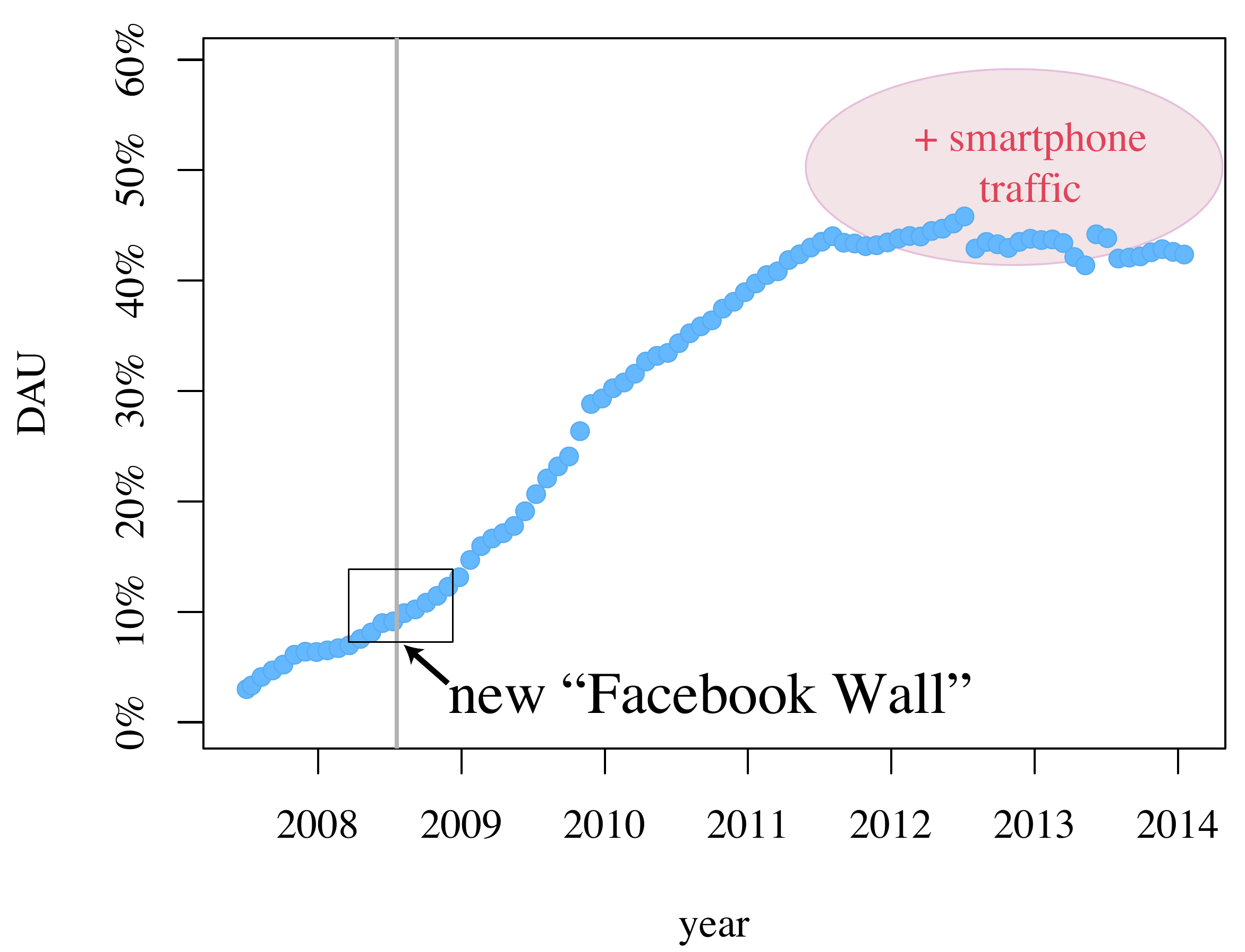}
}\\
\subfloat[][MySpace DAU.\label{f:msd}]{
\includegraphics[width=1.6in,height=1.2in]{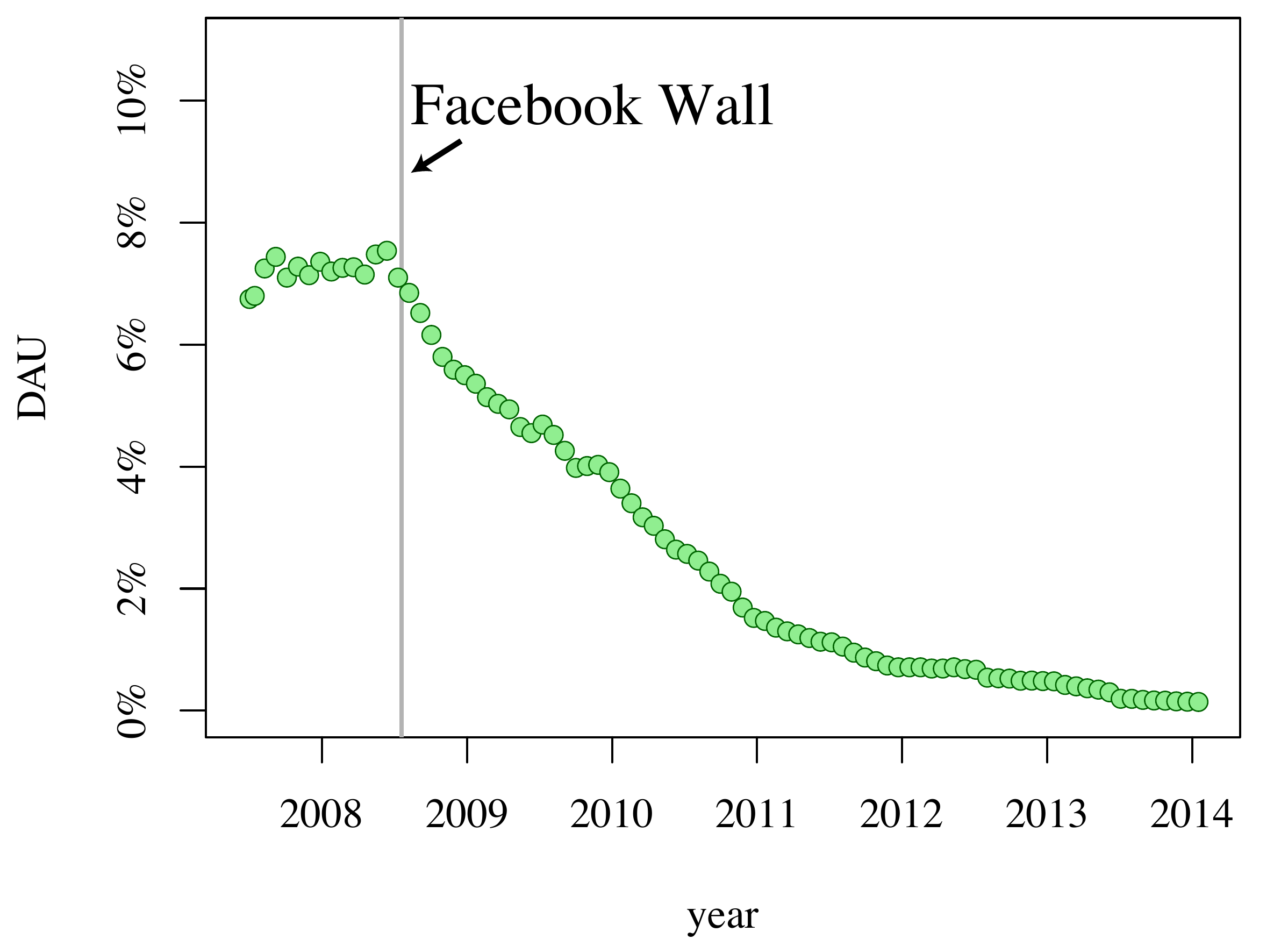}
}
\subfloat[][Hi5  DAU.]{
\includegraphics[width=1.6in,height=1.2in]{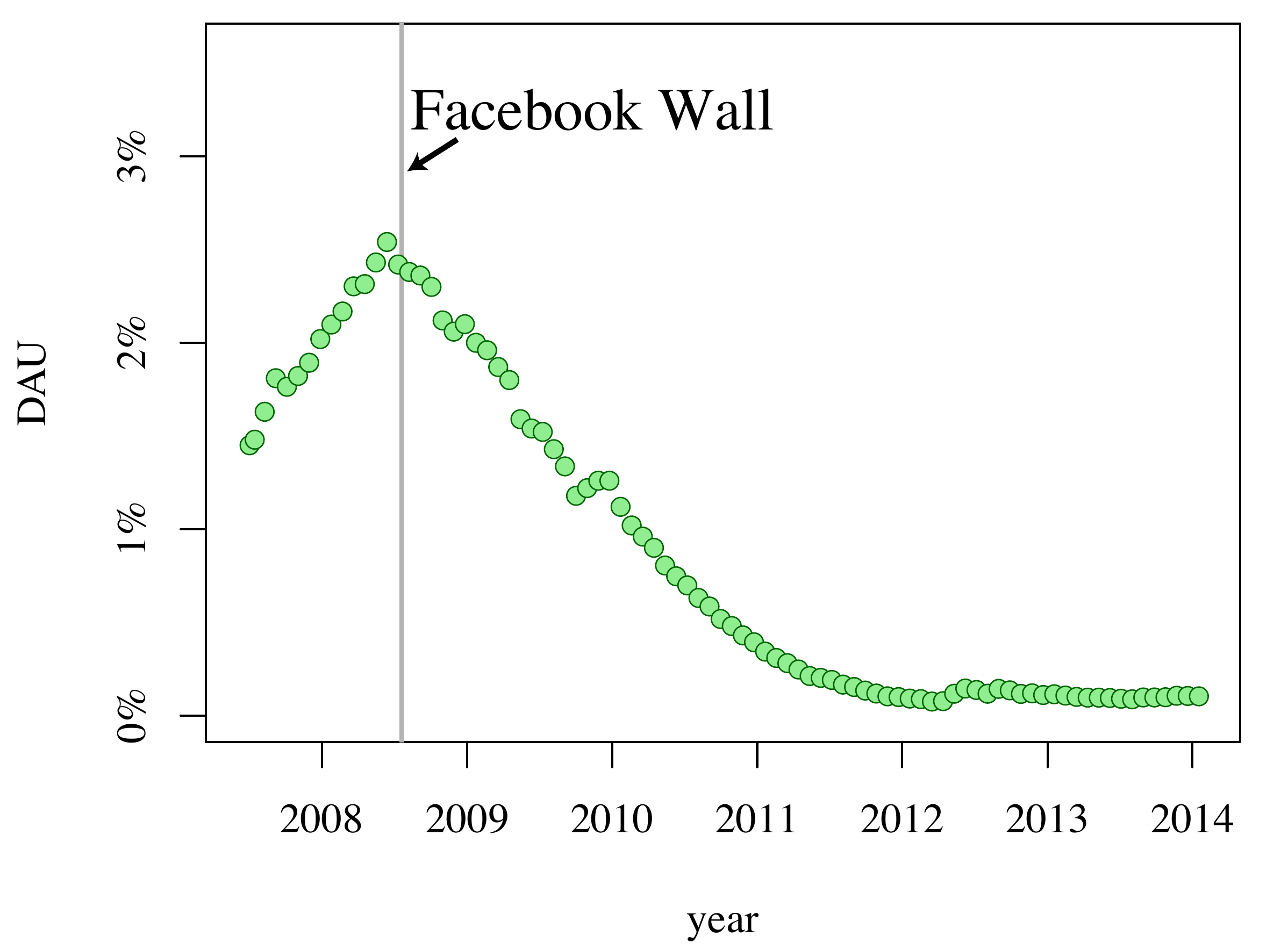}
}
\\
\subfloat[][Friendster DAU.\label{f:frie}]{
\includegraphics[width=1.6in,height=1.2in]{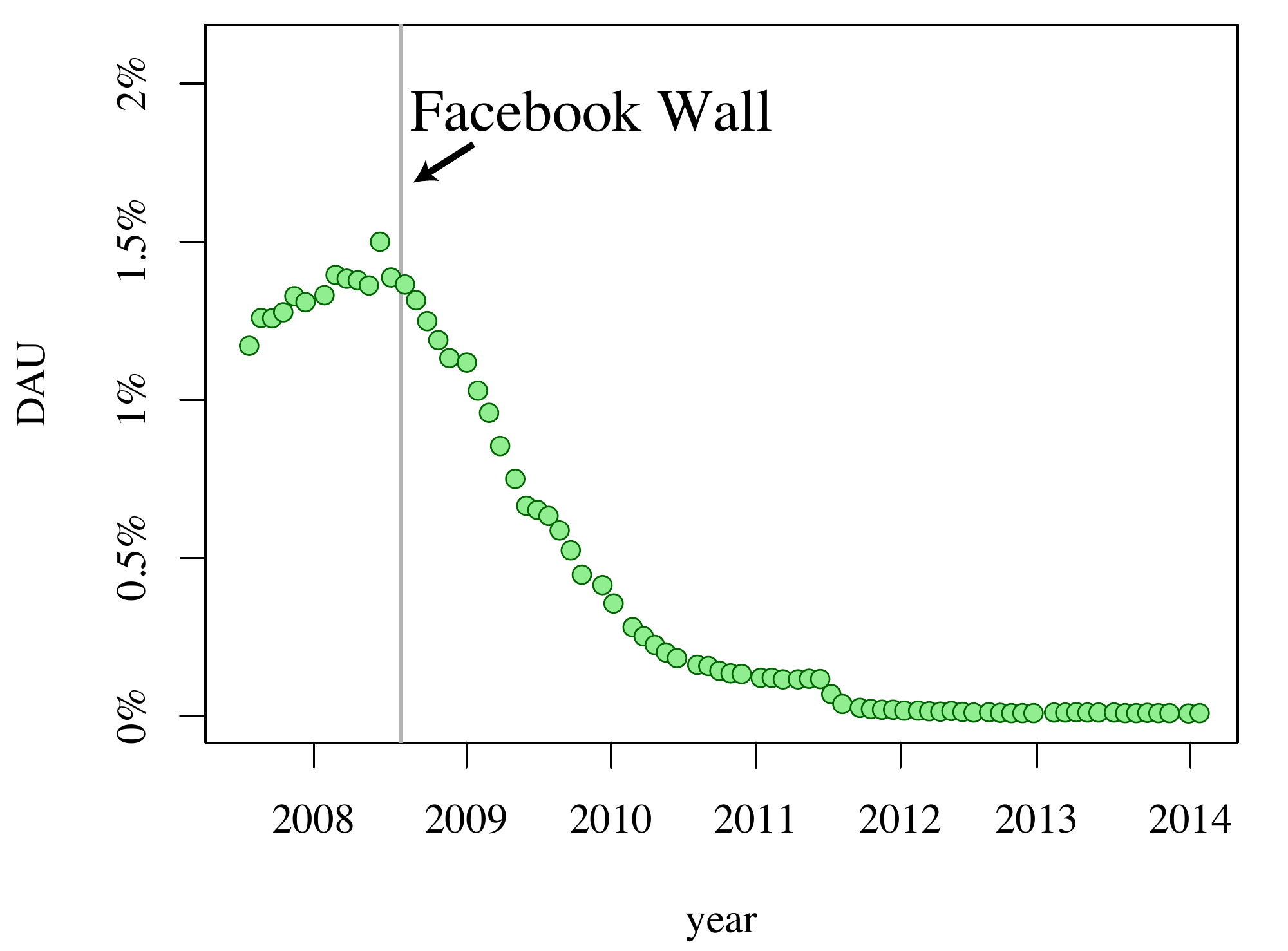}
}
\subfloat[][Multiply DAU.]{
\includegraphics[width=1.6in,height=1.2in]{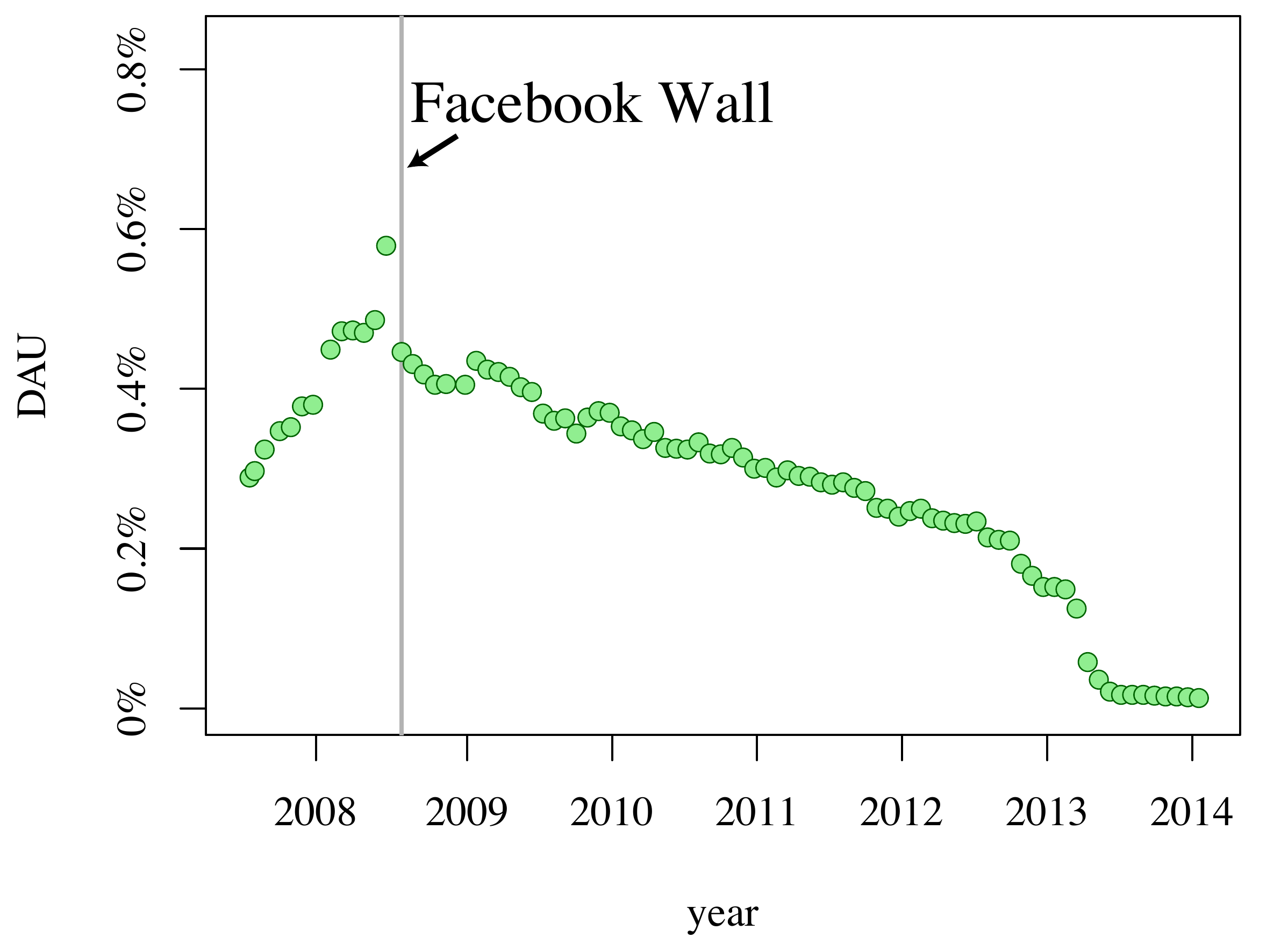}
}
\caption{{\bf (Facebook's Popularity Competition)} DAU/AIP of Facebook, MySpace,  Hi5, Friendester, and Multiply websites from June 2007 to February 2014. Gray vertical lines show the time of the introduction of the ``new Facebook''. \label{f:OSN}}
\vspace{8pt}
\end{figure}

Conversely, a website that suddenly increases its attention consumption may prey on the attention of its competitors, driving them to their (popularity) death; website death can happen as a result of a negative attention-activity feedback loop (less attention$\to$less content$\to$less attention) that is a function of the size of the concurrent adopter population and the amount of attention (time) left to the competitor.
This observation helps explain how a Facebook website change may have  suddenly (July 2008) interfered with the popularity of its competitors after a long period of non-interfering co-existence (see Figure~\ref{f:OSN}).
The July 2008 Facebook user behavior change is documented in Viswanath et al.~\cite{Viswanath:2009en}.
Section~\ref{s:datasets} offers more details about this event.

Last, and perhaps almost as importantly, our model showcases the possibility of predicting popularity trends of competing websites using the widely available DAU time series.
Using training data that includes a few months-worth of DAU data after the competition for attention starts to show a clear signal in the DAU time series, 
our model is able to accurately forecast the popularity of two competing websites nearly  five years into the future.
We also show that our model principles are consistent with detailed user-level activity of 250,000 MySpace users measured between 2004 and 2009.
\vspace{8pt}

\subsection*{Outline}
The outline of this work is as follows.
Section~\ref{s:datasets} introduces the websites and datasets used in this study and presents key DAU patterns in the Facebook, MySpace, Hi5, Friendster and Multiply DAU data. These patterns are used to inform a set of desired model properties that are used to guide the development of our model. 
Section~\ref{s:related} presents an overview of the related work.
Section~\ref{s:model} introduces our model.
Section~\ref{s:pred} fits the model parameters to the data and uses the fitted model to predict the DAU years into the future.
Finally, Section~\ref{s:conclusions} presents our conclusions and future work.

\section{Desired Model Properties}\label{s:datasets}

In what follows we draw key findings from our datasets and the existing literature to inform our model design through a set of required model properties.
The next section, Section~\ref{s:related}, contrasts the existing literature through the lenses of these desired model properties.

\subsection{Datasets}
In this work we use two complementary sources of data.
The first datasets are provided by Amazon's Alexa web analytics company totaling $32$ years of DAU data.
The DAU time series is measured from June 2007 to January 2014 as a fraction of the total Active Internet Population (AIP) of each day.  
We note in passing that Alexa's DAU/AIP measurements of Facebook and other websites may have a {\bf strong U.S.\ and Canada bias}.
A good argument for using the DAU/AIP instead of the raw DAU value is that the DAU/AIP ameliorates seasonal effects such as school breaks and holidays.
But in order to simplify our notation, throughout this work we use DAU to refer to the quantity DAU/AIP. 
As standard practice we smooth out the DAU outliers using a moving median with a 31-day DAU interval centered around each day.
Our second dataset records the activity of 250,000 MySpace subscribers from 2004 to early 2009, collected by Ribeiro et al.~\cite{NetSciCom2010}.

\begin{table*}[!!!th]
\centering
\ra{1.1}
\begin{tabular}{@{}lccccc@{}} 
\toprule 
& \multicolumn{5}{c}{Model Properties} \\ 
\cmidrule(r){2-6} 
\multirow{2}{*}{Model}  & \multirow{2}{*}{Competition} & Attention-Activity  & Attention & Concurrent  & Disjoint Interfering \\ 
 & &  Feedback  & Sharing &  Adoptions &   Adopters \\
\midrule 
Proposed Model & \cmark & \cmark&\cmark &\cmark  &\cmark\\
Ribeiro~\cite{Ribeiro:2014www} &  \xmark & \cmark & \xmark&\xmark  &\xmark \\
Beutel et al.~\cite{beutel2012interacting} & \cmark &\xmark  & \xmark & \cmark  & \xmark \\ 
Viswanath et al.~\cite{Viswanath:2009en} & \cmark &\xmark  & \xmark & \cmark  & \xmark \\ 
Cauwels and Sornette~\cite{MN_Sornette2012}  & \xmark & \xmark & \xmark&\xmark &\xmark \\
Network Effect Adoptions &   \cmark & \xmark   & \xmark& \xmark  &\xmark \\
Diffusion of Innovation & \cmark &  \xmark  & \xmark& \cmark  &\xmark \\
Threshold Adoption &   \cmark & \xmark & \xmark& \xmark  &\xmark \\
\bottomrule 
\end{tabular}
\caption{Model properties matrix\label{t:models}}
\end{table*}

Figure~\ref{f:OSN} shows the DAU time series of Facebook, MySpace, Hi5, Friendester, and Multiply.
It is important to note that Alexa's datasets do not include smartphone traffic.
Even without Facebook's smartphone data its usage reaches an impressive 45\% of the AIP.
According to Facebook's own (unverifiable) records, adding smartphone-only users takes the DAU to 60\% of the U.S.\  and Canada AIP~\cite{DevlinFB}, an extra 15\% DAU of what is reported by Alexa.
Facebook's 60\% DAU is reported to have remained stable in the last couple of years~\cite{DevlinFB}.
We are interested in the first years of the competition between Facebook, MySpace, Hi5, Friendester, and Multiply, a time when smartphone-only usage was likely small.
{\em Thus, the DAU omission of smartphone traffic should not affect our analysis.} 
However, as a reference for our predictions, we include a circle in Figure~\subref*{f:fb} to represent the uncertainty that Facebook's smartphone-only DAU adds to our data.
In what follows we provide a brief overview of the websites analyzed in this work:
\begin{itemize}
\item {\bf myspace.com:} MySpace was founded in 2003 and from 2005 until early 2008 MySpace was the most visited social networking website in the world.
In June 2006 MySpace surpassed Google as the most visited website in the United States. 
But by April 2008 Facebook usage overtook MySpace~\cite{crunchbase}.
\item {\bf facebook.com:} Facebook was founded on February 4, 2004. It was initially limited to students at various other universities but soon it was opened to any individual older than 13. Facebook is the largest online social network in the world today. Recently its IPO raised \$16 billion, making it the third largest in U.S.\ history~\cite{crunchbase}.
\item {\bf hi5.com:} Founded in 2003, Hi5 is an online social network where users can share photos and play games. Today, social games, virtual goods, and other premium content monetizes the website~\cite{crunchbase}.
\item {\bf friendster.com:} Friendster launched in 2002 as one of the first social networking sites. The service allowed users to communicate with other members, share online content and media, discover new events, brands, and hobbies. The site, at its peak, reached tens of millions of registered users according to CrunchBase~\cite{crunchbase}.
\item {\bf multiply.com:} Multiply is a mix between an e-commerce platform and a social networking website, offering sellers a combination of e-commerce and social communications tools. The website ceased operations in May, 2013~\cite{crunchbase}.

\end{itemize}

\subsection{Desired Model Properties}
In what follows we use our data together with other measurements reported in the literature to suggest key properties that we use as guiding principles of our model of website popularity competition.

\pagebreak
\subsubsection{Attention-Activity Feedback Mechanism}
Attention is a scarce resource that must be consumed by websites.
With the captured share of attention a website engages users into content creation, which in turn further captures the attention of other users.
Users also have other interests apart from the website that also compete for their attention.
Our previous success in modeling the DAU times series through the attention-seeking interaction between users of successful and unsuccessful membership-based websites in Ribeiro~\cite{Ribeiro:2014www} showcases the value of this property in popularity models.
And indeed, recent results coming out of Facebook~\cite{backstrom2011center} indicate that the activity of our friends on Facebook incites us to login and become active which, in turn, incites our friends to either become active or stay active.

The attention-activity feedback mechanism may also come about due to the marginal increase in website utility as it gains more active users, an effect known as {\em network effect} or {\em network externality} discussed at length in Farrell and Klemperer~\cite{farrell2007coordination}. 
There are many types of network effect, but the most widely used effect in its purest form can be described in the following {\em path-dependent cumulative return} rule (see Arthur~\cite{Arthur} for more details): higher DAU $\to$ more advertisement revenue $\to$ better website features $\to$ less inactive users (increased DAU).

Extrapolating the above observations to all online social interactions (email, chat, OSN, blog activity) informs the first requirement of our model:
\begin{quote}
{\em The popularity of a website should be modeled by an attention-activity feedback mechanism between all user activities, both inside and outside the website of interest.}
\end{quote}

\subsubsection{Concurrent Adoptions}

A key factor in modeling popularity competition lies in the  interactions of concurrent adopters, users that have accounts in both competing websites.
The size of concurrent adopter population may be affected by factors that prevent concurrent adoptions, as follows: Once an individual joins website $a$ she will be less interested in joining other competing websites, either due to the {\em network effect} (e.g., her friends are all in $a$), adoption cost (e.g., her ``things'' are all in $a$), or because her ``product needs'' are already fulfilled by $a$ (e.g., why join two RSS news aggregators?).
Henceforth, we refer to this effect as the {\em inertia} effect, a force that (at least initially) opposes concurrent adoptions.
The {\em inertia} effect is one of the main arguments in favor of network effect~\cite{Bala:2000do,farrell2007coordination,Garcia:2013ty,Jackson:2005tm,katz1985network,liebowitz1994network,Marsili:2004kf,Montanari:2010cw,Skyrms:2009gu,Snijders:1996ea,Young:2011br} and threshold~\cite{granovetter1978threshold,schelling2006micromotives} adoption models.

However, as it is often the case in social studies, the {\em opposite} explanation seems as plausible: website adopters -- specially the ones that do not adopt the ``leading'' website as with Hi5 and Google+ adopters  -- are more likely to be ``technology enthusiasts'' than the average Internet user and, thus, also more likely to join multiple websites.
The case exists for the existence of concurrent adoptions in the wild, as documented by Goga et al.~\cite{Goga:2013vv}.
We refer to this effect as the {\em momentum} effect, where a user adopting a website signals a higher-than-average likelihood of her adopting its competitor.

In real world scenarios the above opposing forces likely co-exist in a population-level sense through distinct users.
\begin{quote}
{\em A model of popularity competition should include a tunable parameter that covers a large spectrum of net population-level effects  of concurrent adoptions, from strong  inertia to strong momentum.}
\end{quote}

\subsubsection{Non-interfering to Sudden DAU Interference}

In Figure~\ref{f:OSN} we observe a sudden synchronous drop in the popularity of MySpace, Hi5, Friendster, and Multiply by July 2008.
The date coincides with the observation of a significant change in Facebook's user behavior~\cite{Viswanath:2009en}.
Analyzing the activity of 60,000 Facebook users between September 2006 and January 2009, Viswanath et al.~\cite{Viswanath:2009en} observed only one significant Facebook user activity change starting in July 20$^\text{th}$ of 2008, which Viswanath et al.\  points out that coincides with the time that Facebook introduced the beta-test of its ``new Facebook'' design~\cite{Viswanath:2009en}.

Indeed, in July 20, 2008 Facebook introduced the ``new Facebook Wall'' with a radically different content-pushing ({\em news feed}) interface, offered only to ``selected users''~\cite{NewFB1}. By September 2008 these ``selected users'' already amounted to 30 million~\cite{NewFB2}.  
The sudden appearance of this new feature -- which could have been disproportionally adopted by ``tech enthusiasts'' concurrent adopters -- only marginally affects Facebook's DAU (the inset in Figure~\subref*{f:fb} shows an insignificant DAU bump) while it shows a noticeable and seemly lasting impact on MySpace, Hi5, Friendster, and Multiply DAU time series.

\begin{figure}[!thb]
\begin{center}
\includegraphics[width=0.45\textwidth,height=2.2in]{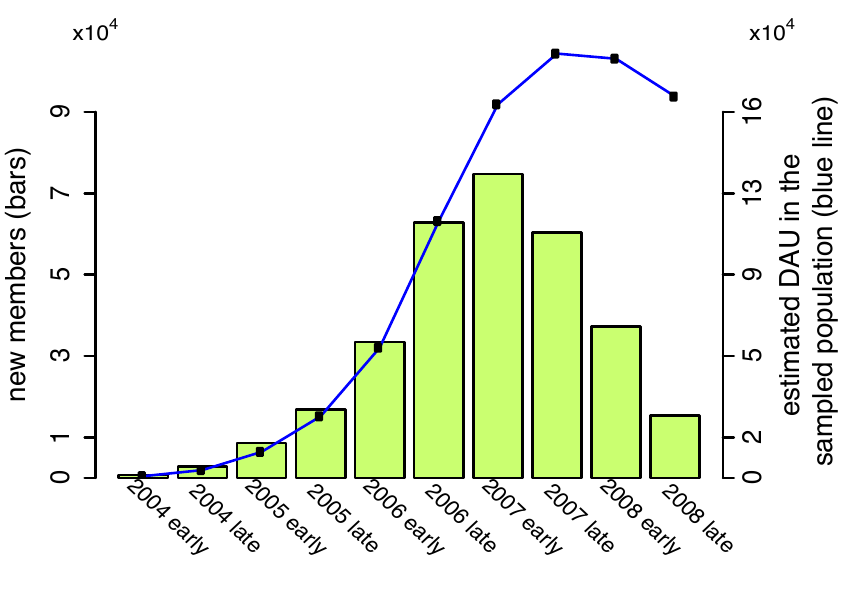}
\end{center}
\vspace{-.15in}
\caption{%
{{\bf MySpace growth $\times$ activity}. 
Number of observed new users (bars) and active users (line) per semester from 2004 to 2008. Note the bell shape in the green bars, a  characteristic of adoption saturations (see Rogers~\cite{rogers1995diffusion}). \label{f:join}%
\vspace{.1in}
}}
\end{figure}

The attention-activity marketplace provides an easy way to assess such sudden changes.
Facebook's growth is accompanied by a growing number of MySpace, Hi5, Friendster, and Multiply adopters that become concurrent adopters.
Figure~\ref{f:OSN} shows that the growing concurrent adopter base does not interfere with MySpace, Hi5, Friendster, and Multiply DAU time series.
This happens because Facebook's attention share does not interfere with concurrent adopter attention to its competitors.
After July 2008 concurrent adopters find themselves spending more time on Facebook, time now taken out of the budget of attention of Facebook's competitors.
A significant enough reduction in attention from concurrent adopters creates a critical mass that affects content creation on these competitors, reducing the attention and content creation (activity) of other users, which in turn further reduces the attention level of concurrent adopters; if this negative attention-activity feedback crosses a particular threshold, the negative feedback loop drives the website to its death.  
The above observation prompts the following model property:
\begin{quote}
{\em Below a given attention budget, a website may consume extra attention from its concurrent adopters without interfering with its competitors. Further attention gains come at the expense of its competitors.}
\end{quote}

\subsubsection{Disjoint Interfering Adopters}

Recent reports indicate that today Facebook user base (penetration) reaches 69\% of the U.S.\ Internet population~\cite{WikipediaFaB}.
In contrast, MySpace, even without Facebook's competition, would in all likelihood never have reached this success.
By mid 2007  -- when Facebook's DAU was at a mere 3\% -- MySpace's DAU was already stable and new adoptions in clear decline showing signs of saturation (for MySpace adoptions see green bars in Figure~\ref{f:join}; for the characteristics of adoption saturation see Rogers~\cite{rogers1995diffusion}).

\begin{figure}[!ht]
\begin{center}
\includegraphics[width=3in,height=1.6in]{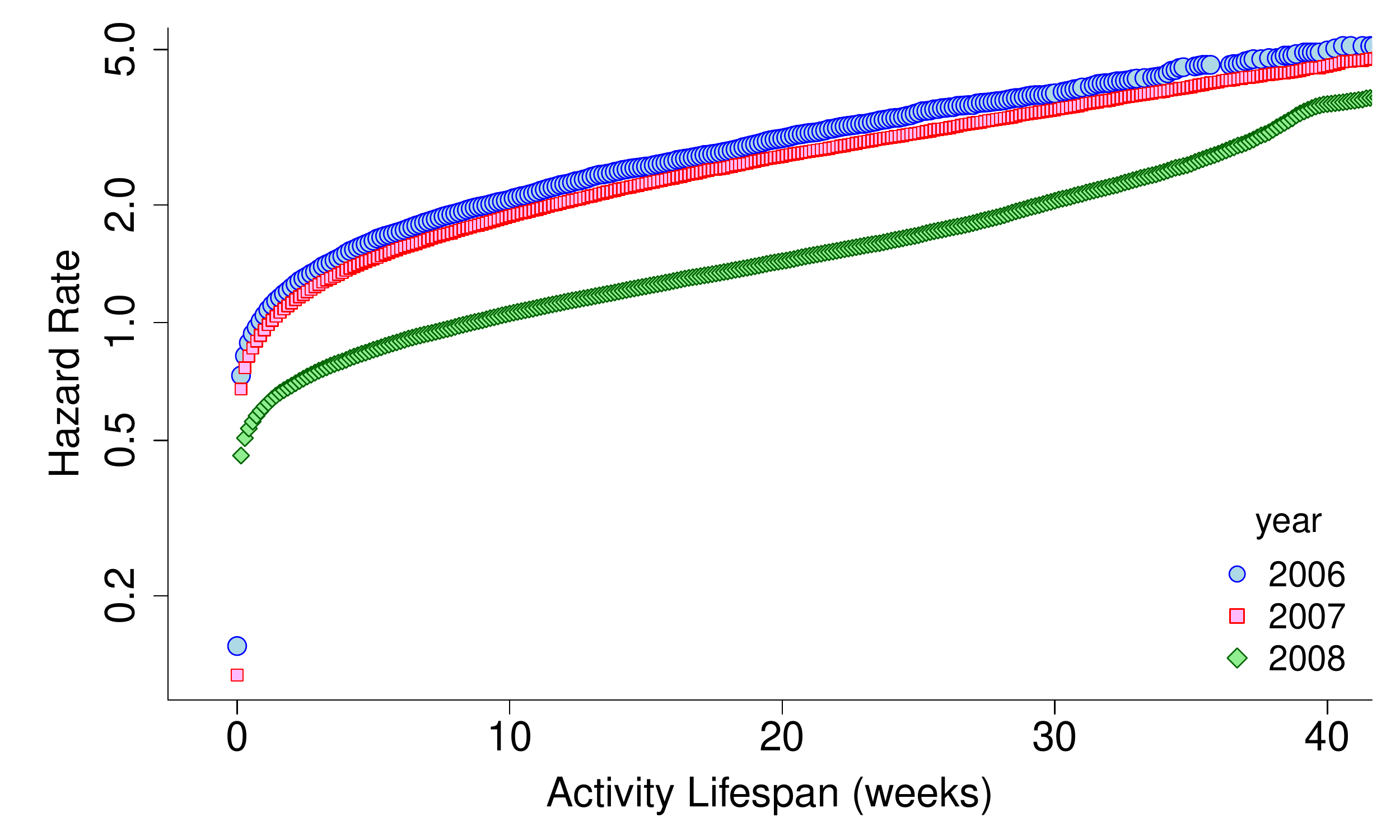}
\end{center}
\caption{
MySpace hazard rate estimates for new subscribers starting in 2006, 2007 and 2008, respectively. Estimates from their first year of activity.\label{f:hz}%
}
\end{figure}

This saturation is unlikely to be due to MySpace's competition for users with Facebook. 
The 2007 ``drop-out'' (hazard) rates of MySpace's new adopters were identical to that of 2006 as shown by the Kepler-Meier hazard rate estimates in Figure~\ref{f:hz}. 
The use of the Kepler-Meier estimator is needed as our MySpace user activity data is right censored (collection stopped by January 2009).
It is only by 2008 that new adopters show different hazard rates for MySpace users signifying that these adopters were in average more committed to MySpace than 2006 and 2007 new adopters. 
Thus, it is unlikely that even if MySpace was allowed to take its course without any competition it would not have reached Facebook's 69\% U.S.\ penetration.
The same probably can be said about Hi5, Friendster, and Multiply although we do not have user-level data for these websites.
The reason behind the opposition to adopt a website -- whether by principle, lack of features of broad appeal, or other factors -- are transparent to the model. 
It is only important to model that such opposition exists.

It is as important, however, to note that a user only interested in one website (say, Facebook) may still indirectly affect the DAU of that website's competitors (say, MySpace and Hi5).
To illustrate, consider Facebook users that are ``opposed'' to joining MySpace.
The activity of these users help create content on Facebook, which in turn attracts the attention of concurrent adopters. 
Thus, the activity of ``Facebook-only'' users can indirectly affect the activity of ``MySpace'' users.
\begin{quote}
{\em Hence, a model of popularity competition should include disjoint populations of adopters that, however, affect each other through the attention and activity of concurrent adopters.}
\end{quote}


\section{Related Work} \label{s:related}
Adoption models describe phenomena as diverse individuals deciding to adopt a new technology~\cite{bass2004comments,farrell2007coordination,fisher1972simple,geroski2000models,katz1985network,liebowitz1994network,Prakash:2012cr,Ribeiro:2014www} or individuals adopting a new health habit~\cite{Christakis:2007fy,centola2010spread}.
These models have deeply influenced the study of online social network growth~\cite{backstrom2006group,centola2010spread,Jin:2001hi,kairam2012life,leskovec2008microscopic,Leskovec07,ugander2012structural}.
In the literature the work that is closest to ours is that of Prakash et al.~\cite{Prakash:2012cr}.
Prakash et al.\  models the popularity of two competing products (e.g., smartphones) using  a generalization of the diffusion of innovation model (detailed below). 
The model in Prakash et al.\ does not capture users finite attention or their inertia/momentum in concurrent adoptions.
Moreover, the model in Prakash et al.\  does not consider disjoint interfering adopters.
Beutel et al.~\cite{beutel2012interacting} extends the model of Prakash et al.\   to include interfering adopters.
Table~\ref{t:models} compares the models in the literature against the desired model properties specified in Section~\ref{s:datasets}.

Our model is a generalization of the our previous single-website model in Ribeiro~\cite{Ribeiro:2014www}, where the popularity of \\ membership-based websites is modeled using a population-level reaction-diffusion-decay model.
Our new model significantly generalizes that model by explicitly modeling user's attention, attention sharing, website competition, and disjoint adoptions in a marketplace where users must share their finite attention.
Note that in our generalization we make do without the spontaneous exponential decay required in Ribeiro~\cite{Ribeiro:2014www} by modeling the attention grabbing influence of other online activities.

Another work closely related to ours is that of Cauwels and Sornette~\cite{MN_Sornette2012} which focuses on describing the evolution of the Facebook DAU.
Cauwels and Sornette, however, is incomplete in the sense that its time-series analysis is tailored towards Facebook's success and, thus, cannot capture DAU decays.
The work of Liu et al.~\cite{Liu:2013vb} models the popularity of applications inside online social networks such as Facebook.
More generally, regarding product adoptions can be classified as: (a) Network effect adoption models (a.k.a.\ network externality models)~\cite{Bala:2000do,farrell2007coordination,Garcia:2013ty,Jackson:2005tm,katz1985network,liebowitz1994network,Marsili:2004kf,Montanari:2010cw,Skyrms:2009gu,Snijders:1996ea,Young:2011br}, where individual rationality and adoption costs and utilities are modeled in a game-theoretic framework; { (b) Threshold adoption models}~\cite{granovetter1978threshold,schelling2006micromotives}, where an individual adopts if enough of his or her friends are adopters; { (c) Diffusion of innovation models}~\cite{bass2004comments,fisher1972simple,geroski2000models,Prakash:2012cr}, where adopters influence others to adopt through word-of-mouth; In the absence of fine-grained individual-level data these models provide demand forecasting at the aggregate (population) level; and finally { (d) Adoption models from influence and network structure}~\cite{ugander2012structural,centola2010spread,backstrom2006group,Jin:2001hi,leskovec2008microscopic,Leskovec07,kairam2012life,Kumar:2010ek}, where an individual adoptions depends not only on whether his or her friends adopt but also on how these friends are connected among themselves.
A variety of works also consider the relationship between community growth inside an online social network (OSN) websites and their network structure~\cite{backstrom2006group,leskovec2008microscopic,kairam2012life}.
These studies, however, focus on (i) the growth of communities inside the OSN (not the growth of the OSN itself) and (ii) the role of network structure disregarding whether the community is alive (active) or dead (inactive). 
A more thorough review can be found in Ribeiro~\cite{Ribeiro:2014www}.

\section{Model}\label{s:model}

Compartmental models of  interacting populations have been successfully applied in mathematical biology~\cite{murray2002mathematical} and social systems~\cite{Colizza:2007ik,Colizza:2008fk}.
Our model considers a large segmented user population that interact through catalytic reactions and media \& marketing diffusions.
Reaction and diffusion processes find applications in chemistry, physics, and applied mathematics~\cite{Colizza:2007ik,Colizza:2008fk,murray2002mathematical,vanWijland:1998jl,Kampen:1981vs}.
{\em We choose to avoid stochastic models (which would allow us to give confidence intervals to our predictions) 
because very little is known about the stochastic behavior of the dynamics between inactive and active members of websites and between the latter and non-members}.

Our model can be described as follows.
The user state is a tuple $(W_a,W_b) \in \{\emptyset,U,A,I\}^2$, where $W_a$ and $W_b$ represent the state of the user with respect to competing websites $a$ and $b$, respectively. 
For each website a user can be in one of four states: $\emptyset$ is a permanent state signifying that the user will never adopt the website; all remaining three states are likely transient: $U$: the user is willing to join but she is still unaware of the website; $A$: the user is an active member of the website, and finally $I$: the user is an inactive member of the website.

\subsection{Catalytic Reactions \& Marketing Diffusion}\label{s:m}
We use the following notation to mark interactions between users in distinct populations.
Let $S_{(W_a,W_b)}(t) \in (0,1)$ be the fraction of the active Internet population at state $(W_a,W_b) \in \{\emptyset,U,A,I\}^2$ at time $t$.
We use the notation
\[
S_{(W_a,W_b)}  \xrightarrow{ \upsilon  \times S_{(W^\prime_a,W^\prime_b)} }  S_{(W^\dagger_a,W^\dagger_b)}
\]
to denote a population of users in state $(W^\prime_a,W^\prime_b)$ acting as catalysts of users in state $(W_a,W_b)$,
inciting them to state $(W^\dagger_a,W^\dagger_b)$ in the next $dt$ time step, with probability $\upsilon dt$, $\upsilon \in \mathbb{R}^+$.
The above notation directly translates into the differential equations:
\begin{align*}
\frac{dS_{(W^\dagger_a,W^\dagger_b)}(t) }{dt} &= \dots + \upsilon S_{(W^\prime_a,W^\prime_b)}(t) S_{(W_a,W_b)}(t) \,, \\
\frac{dS_{(W_a,W_b)}(t) }{dt} &= \dots - \upsilon S_{(W^\prime_a,W^\prime_b)}(t) S_{(W_a,W_b)}(t)  \, ,
\end{align*}
where ``$\dots$'' represents the contributions of other reactions and diffusions that flow into the same state.
We also use of other two important definitions: $(W_a,\star)$ represents users with state $W_a$ on website $a$ and any state on website $b$. We also use $\overline{S}_{(W_a,W_b)}(t)$ to denote the fraction of the Internet population that is not in state $(W_a,W_b)$ at time $t$.
{\bf The DAU of website $a$ is given by $$S_{(A,\star)}(t) = \sum_{k \in \{\emptyset, U, A, I\}} S_{(A,k)}(t)$$ and website's $b$ DAU is $$S_{(\star,A)}(t)= \sum_{k \in \{\emptyset, U, A, I\}} S_{(k,A)}(t).$$}
\subsubsection{Disjoint Population Dynamics}\label{s:dj}
The equations describing the dynamics of users of  website $a$ that will never join website $b$ are as follows (a symmetric set of equations model website $b$ users that will never join website $a$).
The first {\bf reaction}
\[
S_{(U,\emptyset)}   \xrightarrow{\gamma_a S_{(A,\star)} }   S_{(A,\emptyset)} \quad \textsf{\{word-of-mouth\}}
\]
describes the catalytic reaction at rate $\gamma_a S_{(A,\star)}$ that happens when an active member of website $a$ influences a unaware users $(U,\emptyset)$ to join website $a$, which can happen either through word-of-mouth or because 
of increased utility (e.g., {\em network effects}),  two widely known phenomena in the specialized literature~\cite{bass2004comments,farrell2007coordination,rogers1995diffusion}.
Unaware users can also join website $a$ through media \& marketing campaign {\bf diffusions}
\[
S_{(U,\emptyset)}    \xrightarrow{\lambda_a }  S_{(A,\emptyset)} \quad \textsf{\{marketing\}}  \, .
\]
The remaining catalytic {\bf reactions} are
\[
S_{(I,\emptyset)}  \xrightarrow{\alpha_a S_{(A,\star)}} S_{(A,\emptyset)} \quad \textsf{\{website activity\}} \, ,
\]
describing the population-level influence that the content created by active users of website $a$ exert on $a$'s inactive users, prodding them into activity (for more details on these dynamics see Ribeiro~\cite{Ribeiro:2014www}); and finally
\[
S_{(A,\emptyset)}  \xrightarrow{\beta_a \overline{S}_{(A, \star)}} S_{(I,\emptyset)} \quad \textsf{\{external activity\}} \: ,
\]
modeling the influence of people doing things other than spending time on website $a$ exert on website $a$ users to also do something else.

\subsubsection{Joint Unaware Population Dynamics}
We now apply the same mechanisms used above to describe the dynamics of website $a$ users that are willing to join website $b$ but 
are still unaware of website $b$. 
In what follows we only present website $a$'s equations; the symmetric corresponding set of equations should be used to describe the dynamics from website's $b$ point of view.
Users unaware of website $a$ join through word-of-mouth and media \& marketing {\bf diffusions}:
\begin{align*}
S_{(U,U)}  &  \xrightarrow{\lambda_a }  S_{(A,U)} \quad \textsf{\{marketing\}} \,,\\
S_{(U,U)}  & \xrightarrow{\gamma_a S_{(A,\star)} }  S_{(A,U)}   \quad \textsf{\{word-of-mouth\}} \, .
\end{align*}
The forces that users on website $a$ and people outside website $a$ exert on each other manifest in the following catalytic {\bf reactions}:
\begin{align*}
S_{(I,U)} &  \xrightarrow{\alpha_a S_{(A,\star)}} S_{(A,U)}  \quad \textsf{\{website activity\}}\,, \\
S_{(A,U)}  & \xrightarrow{\beta_a \overline{S}_{(A, \star)}} S_{(I,U)} \quad \textsf{\{external activity\}} \, .
\end{align*}
Up until now websites $a$ and $b$ do not interfere with each other.
In what follows we consider the dynamic of concurrent adopters.
But first users must become concurrent adopters.
A user of website $a$ becomes a concurrent adopter of website $b$ through the following catalytic {\bf reactions} and {\bf diffusions}:
\begin{align*}
S_{(A,U)} &\xrightarrow{\zeta_a  \lambda_b } S_{(A,A)} \quad \textsf{\{marketing+inertia|momentum\}} \,,\\
S_{(A,U)} &\xrightarrow{\zeta_a \gamma_b S_{(\star, A)}} S_{(A,A)} \: \textsf{\small\{word-of-mouth+inertia|momentum\}}\,, \\
S_{(I,U)} &\xrightarrow{\zeta_a  \lambda_b } S_{(I,A)} \quad \textsf{\{marketing+inertia|momentum\}} \,,\\
S_{(I,U)} &\xrightarrow{\zeta_a \gamma_b S_{(\star, A)}} S_{(I,A)} \: \textsf{\small\{word-of-mouth+inertia|momentum\}}  \, ,
\end{align*}
where $\zeta_a \in \mathbb{R}^+$ is a parameter that covers the spectrum of net population-level effects from {\bf inertia} for $ \zeta_a < 1$ to {\bf momentum} for $\zeta_a > 1$.
In order to reduce the model complexity and also model the relative interest generated by the websites over time, we can further
reduce the parameter space of our model $\zeta_a(t) = \zeta S_{(A,\star)}(t)/(S_{(A,\star)}(t)+S_{(\star,A)}(t))$ and $\zeta_b = \zeta S_{(\star,A)}(t)/(S_{(A,\star)}(t)+S_{(\star,A)}(t))$, where $\zeta \in \mathbb{R}^+$.
In what follows we cover the dynamics of concurrent adopters.
\subsubsection{Concurrent Adopters Dynamics} \label{s:ca}
In our current attention-activity marketplace model concurrent adopters do not interfere with each other's inactive$\to$active dynamics.
This is because spending more time (attention) on website $a$ is assumed not to make the activity of other users (say, on website $b$) seem less interesting.
Using the fact that Facebook suffered nearly no lasting DAU effect around the July 2008 website redesign as a guiding principle (as shown by the tiny DAU bump in the inset of Figure~\subref*{f:fb}), we will assume that the extra time spent on website $a$ does not increase the user activity (attention-grabbing activity) rate $\alpha_a$, that is, $\alpha_a$ remains unchanged in the following catalytic {\bf reactions}:
\begin{align*}
S_{(I,A)}  &  \xrightarrow{\alpha_a S_{(A,\star)}}  S_{(A,A)} \quad \textsf{\{website $a$ activity\}} \,,\\
S_{(A,I)}  &  \xrightarrow{\alpha_b S_{(\star, A)}}  S_{(A,A)} \quad \textsf{\{website $b$ activity\}}\,, \\
S_{{(I,I)}}  &  \xrightarrow{\alpha_a  S_{(A,\star)}}  S_{(A,I)} \quad \textsf{\{website $a$ activity\}} \,,\\
S_{{(I,I)}}  &  \xrightarrow{\alpha_b  S_{(\star, A)}}  S_{(I, A)}  \quad \textsf{\{website $b$ activity\}}\,. 
\end{align*}
Interestingly, according to our framework an increase in $\alpha_a$ happens only when a new ``user activity'' feature is added, which our model predicts will cause an abrupt permanent change in the DAU slope.
For instance, in February 2009 Facebook introduced the ``Like'' feature~\cite{WikipediaFB} and, indeed, in Figure~\subref*{f:fb} we observe a small but sustainable sharp jump in the DAU time series.
Similarly, in September 2009 Facebook introduced the ``tagging'' feature~\cite{WikipediaFB} and, again, we observe another DAU sustainable sharp jump followed by a slope change.
In these scenarios $\alpha_a$ becomes $\alpha_a(t)$, a right-continuous step function that sharply changes after a new ``user activity'' feature is added.
In order to keep model complexity down in our experiments and because these unpredictable changes are not of interest to out forecast,  we consider $\alpha_a$ as a constant in the results presented in Section~\ref{s:pred}.

\paragraph{Attention Sharing of Concurrent Adopters}
In what follows we model the attention sharing behavior of users.
In the attention-activity marketplace the concurrent adopters are responsible for driving otherwise self-sustaining websites (see Ribeiro~\cite{Ribeiro:2014www} for a precise definition of website self-sustainability) to their ``unnatural'' death.
In our model the average time a user spends on the website is latent, as we are modeling the DAU.
However, we can model its effect on the parameters of our model.
Let $B$ be the average time budget of time that a user is willing to spend at online social interactions.
Let $B_a$, $B_b$, and $B_o$ be the times that the user spends at websites $a$, $b$, and at other activities $o$ s.t.\ $B_o = B - B_a - B_b$.
If at time $t_0$ $B_a$ sharply increases by $\Delta_a$ and $B_{o}$ sharply decreases by $\Delta_a$ then $B_b$ remains constant. 
In this scenario the DAU of website $b$ remains unchanged and websites $a$ and $b$ do not interfere with each other.

However, if the sharp increase in $B_a$ by $\Delta_a$ is met with a decrease $\Delta_b$ in $B_b$, i.e., users are not willing to further compromise $B_{o}$, then 
$\beta_b$ also abruptly changes to follow the abrupt change in $B_b$ as $\beta_b (1 + \Delta_b/B_b)$.
Define $\eta_b^\prime := \Delta_b/B_b$;
using the Heaviside step function, $H(t_0)$, at time $t_0$ yields the catalytic {\bf reactions} that model the attention sharing behavior of concurrent adopters:
\begin{align*}
S_{(A,A)}  &  \xrightarrow{\beta_a  \overline{S}_{(A, \star)}}  S_{(I,A)} \quad \textsf{\{external activity\}} \,, \\
S_{(A,I)}   &  \xrightarrow{\beta_a \overline{S}_{(A, \star)}}  S_{{(I,I)}}  \quad \textsf{\{external activity\}}\,,\\
S_{(A,A)}  &  \xrightarrow{(1 + H(t_0)\eta_b) \beta_b   \overline{S}_{(\star, A)}}  S_{(A,I)} \quad \textsf{\{external activity\}}\,, \\
S_{(I,A)}   &  \xrightarrow{(1 + H(t_0)\eta_b) \beta_b \overline{S}_{(A, \star)}}  S_{{(I,I)}} \quad \textsf{\{external activity\}}\, .
\end{align*}

\section{DAU Model Fit}
In this section we briefly introduce the challenges of learning the parameters of our model from the DAU data.
The DAU time series only provides information about $S_{(A,\star)}$ and $S_{(\star,A)}$.
The parameters introduced in Section~\ref{s:m}  need to be estimated together with the population compartment fractions. The latter is what mathematical biologists call the {\em carrying capacity} of each of our four compartments, represented by the population fractions: (1) users ``opposed'' to both websites $a$ and $b$: $C_{00} := S_{(\emptyset,\emptyset)}$; (2) the users opposed to website $b$: $C_{10} := \sum_{k \in \{U,A,I\}} S_{(k,\emptyset)}$; (3) users opposed to website $a$: $C_{01} := \sum_{k \in \{U,A,I\}} S_{(\emptyset, k)}$; and (4) the fraction of concurrent adopters 
$$
C_{11} := \sum_{k_a \in \{U,A,I\}} \sum_{k_b \in \{U,A,I\}} S_{(k_a,k_b)},
$$
such that $C_{00} + C_{10} + C_{01} + C_{11} = 1$.

Note that because the DAU time series starts June 18, 2007 and not when the websites were created, we also need to parametrize the unobservable quantities $S_{(I,\star)}(t_{-1})$ and $S_{(\star,I)}(t_{-1})$, where $t_{-1} = $``June 18, 2007''.
For the remaining quantities that need to be initialized with values greater than zero due to the $t_{-1}$ start, e.g., $S_{(A,A)}(t_{-1})$, we use independence assumptions, e.g., $S_{(A,A)}(t_{-1}) = S_{(A,\star)}(t_{-1}) S_{(\star,A)}(t_{-1})$.
We fit the model parameters to the DAU data using the Levenberg-Marquardt algorithm~\cite{levenberg44}.
Our results in Section~\ref{s:pred} show the model fit using the first two years of DAU data to train the model, the following four months for model selection,  and the remaining years as holdout data to evaluate the model predictions.

The Levenberg-Marquardt algorithm only finds a locally optimal solution starting from an initial parameter guess.
Hence, the initial guess may significantly influence the output of the algorithm.
Due to the large number of parameters of our model we run the Levenberg-Marquardt algorithm with multiple initial parameter guesses,
choosing the fitted parameters that best fit our model selection data.
In order further reduce the number of parameters to be learned, we tried the options of learning $C_{11}$ or setting $C_{11} = \min(C_{10},C_{01})$. While these two options give similar results, the option $C_{11}$ requires at least ten times the number of initializations and, in the end, $C_{11} = \min(C_{10},C_{01})$ in all examples we tried.
Therefore, in the scenarios presented in Section~\ref{s:pred} we set $C_{11} = \min(C_{10},C_{01})$ to speed up computations.

\section{Results}\label{s:pred}
In this section we briefly introduce our results.
In all of our results an extra four months of DAU data is used to select the best parameter fit that does not overfit the data (model selection phase).
The fitted models and their predictions show great agreement with the data.
But the learned parameter should be interpreted carefully given that most differential equations in our model are quadratic (the ones with $\zeta$ are cubic) and the model has a multitude of parameters and latent variables.
The only observable quantities (the DAUs) are the aggregates $S_{(A,\star)}$ and $S_{(\star,A)}$, the DAU data is left-censored (our DAU time series starts mid 2007 when, for instance, MySpace was already four years old), only two years of DAU data are used to fit the model, and the maximum DAU of Facebook in the training set is 20\%.

Nevertheless, the learned parameter seem to offer interesting insights into the popularity growth of Facebook and the death of its competitors.
As expected, the model predicts {\em momentum} in concurrent adoptions for all websites, that is, users of MySpace, Hi5, Friendster, and Multiply were more likely than the average user to adopt Facebook.

\begin{figure}[!!htb]
\begin{center}
\includegraphics[width=3.2in,height=2.5in]{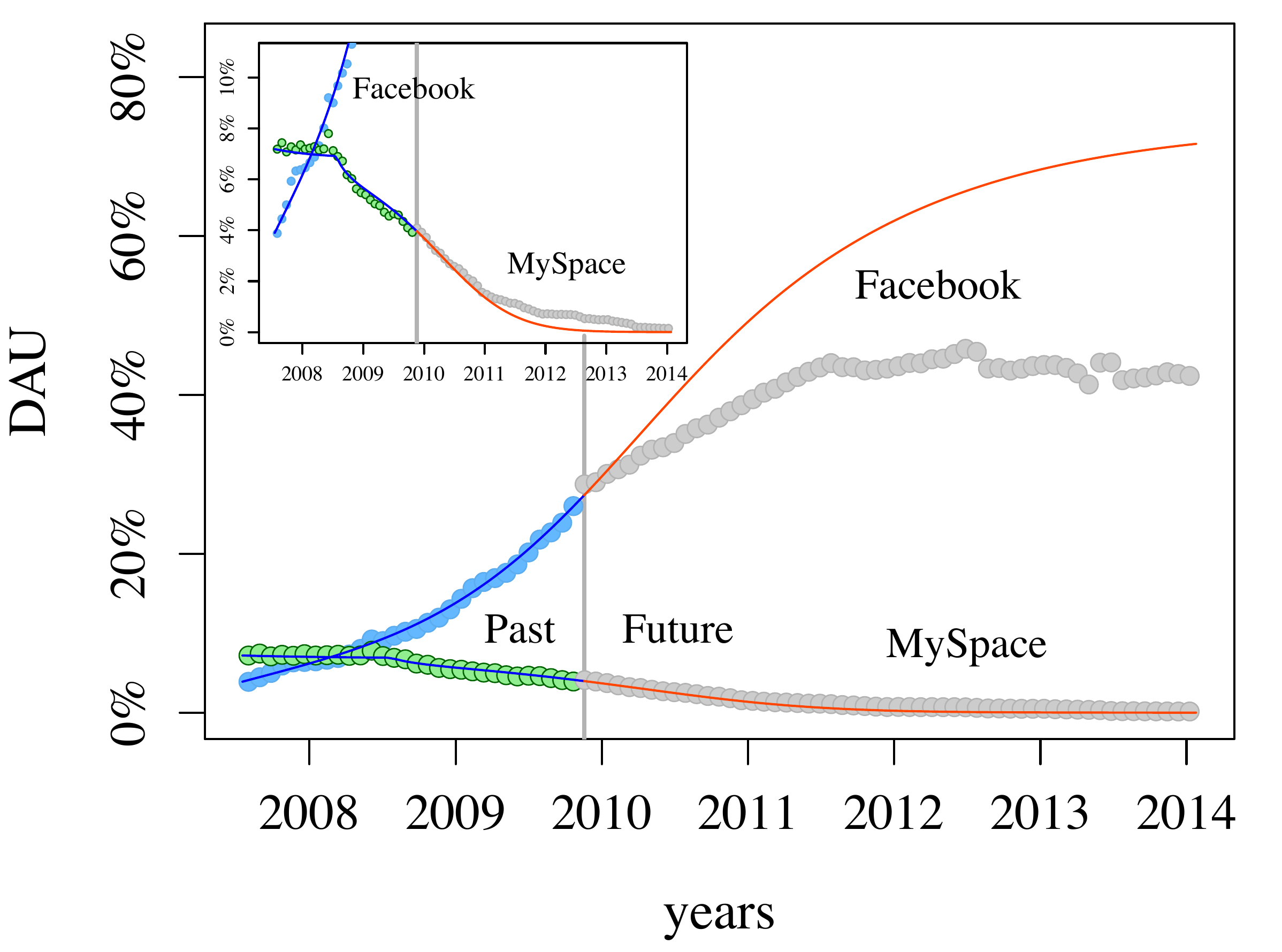}
\end{center}
\vspace{-8pt}
\caption{Model fit and predictions for the competition Facebook v.s.\ MySpace.\label{f:MSR}}
\end{figure}

\paragraph{Facebook v.s.\ MySpace}
Figure~\ref{f:MSR} shows the results of the model fit using the first two years (24 months) of the DAU time series of the competition between Facebook and MySpace.
These $28$ months (24 months for training and $4$ months for the model selection phase) are shown as blue points in the plot.
The gray vertical line separates the training \& model selection data from the 
remaining 50 months (4.1 years) of DAU data used to test our predictions, shown in the plot as gray points.
The model shows great agreement with the data, both in the training (blue line) and prediction (red line) phases.
The inset in Figure~\ref{f:MSR} gives a closer look at the MySpace results.

The model -- which was trained with Facebook's peaking at only $20\%$ DAU -- estimates the total (unaware + active + inactive) Facebook population at $C_{10}+C_{11} = 77\%$ and MySpace's population at $C_{01}+C_{11} = 22\%$; the model also estimates that Facebook user base grew largely due to word-of-mouth and that 
the July 20, 2008 change in Facebook reduced the average time spent on MySpace by 88\%.
The model fit also estimates $\zeta = 4$, that is, by 2008 MySpace users were about twice as likely to join Facebook than the average Facebook adopter (recall that $\zeta_a(t) = \zeta S_{(\star,A)}(t)/(S_{(\star,A)}(t)+S_{(A,\star)})(t)$).

In light of our findings, it seems that MySpace's documented  ``white flight'' and ``teen disengagement'' in 2007~\cite{Coll_Boyd2011} -- often anecdotally cited by the lay press as the primary reason of MySpace's death -- may have had only a marginal role in MySpace's demise.
While in hindsight many social causes could explain MySpace's demise, they would not explain why Hi5, Friendster, and Multiply simultaneously suffered the same effect.
Moreover, by 2007 MySpace's DAU was stable -- and our model predicts it would have remained stable in the absence of Facebook's change --, thus, making the ``white flight'' and ``teen disengagement'' 2007 hypothesis rather unlikely as the main cause of MySpace's death. 

\begin{figure}[!!htb]
\begin{center}
\includegraphics[width=3.2in,height=2.5in]{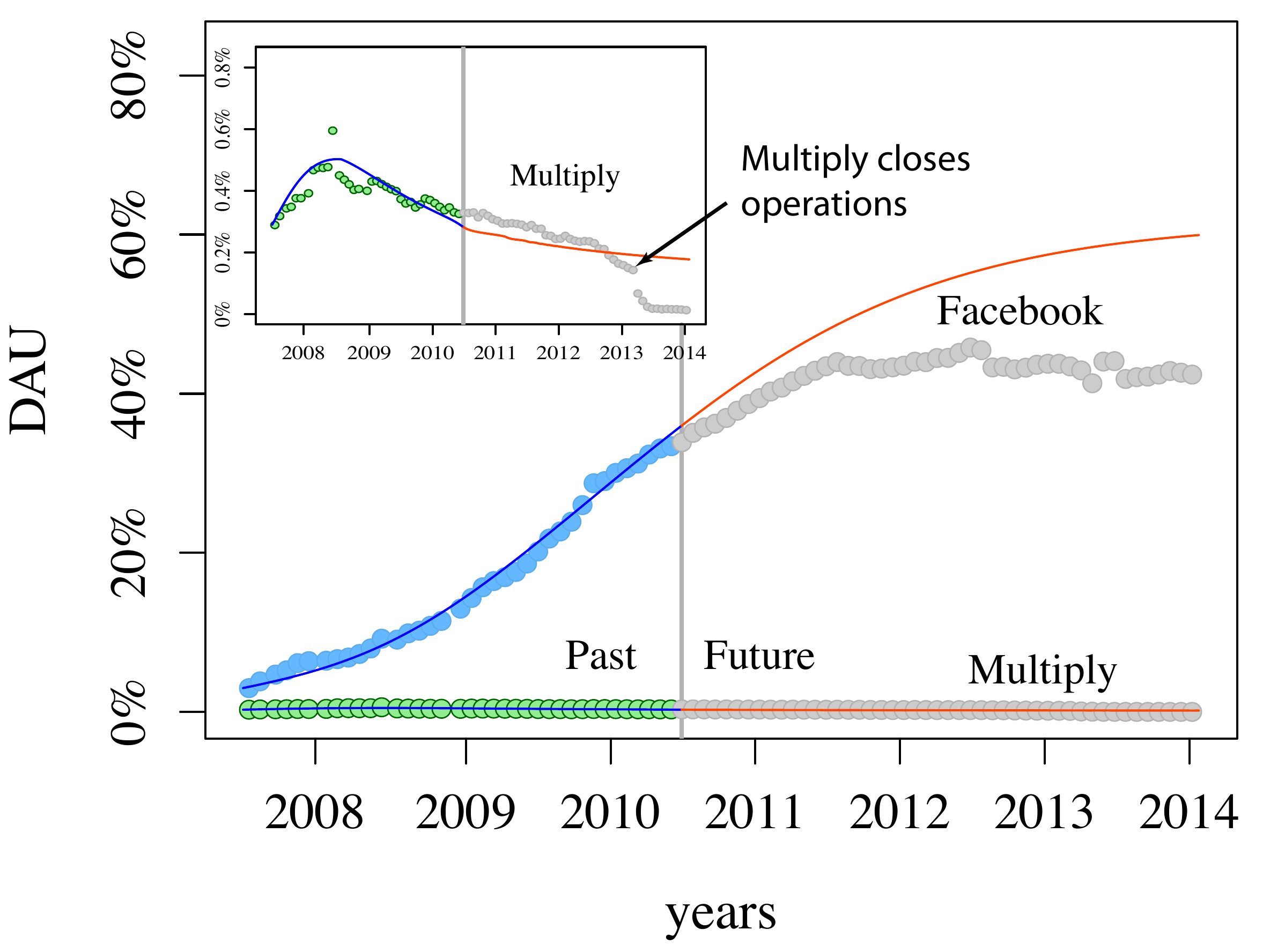}
\end{center}
\vspace{-8pt}
\caption{Model fit and predictions for the competition Facebook v.s.\ Multiply.\label{f:mul}}
\vspace{8pt}
\end{figure}

\paragraph{Facebook v.s.\ Multiply}
Figure~\ref{f:mul} shows the results of the model fit using the first two and a half years (31 months) of the DAU time series in the competition between Facebook and Multiply.
The larger training data was required to achieve a better quality fit of the parameters (harder to learn on Multiply).
Unlike MySpace, Hi5, and Friendster, the model now predicts that Multiply survives Facebook's ``attention raid''.
The model fit shows great agreement with the data up until May, 2013 when Multiply officially closed operations.
The model estimates the total  (unaware + active + inactive) Facebook population at $C_{10}+C_{11} = 64\%$ and Multiply's population at $C_{01}+C_{11} = 0.9\%$; the model estimates that both Facebook and Multiply user base grew largely due to word-of-mouth.
The model also estimates that immediately after the July 20, 2008 event the average time spent on Multiply decreases ``only''  by 24\%, which is why Multiply is projected to survive.
The model fit also estimates $\zeta = 7.5$, that is, Multiply users were over seven times as likely to join Facebook than the average Facebook adopter.

\begin{figure}[!!htb]
\begin{center}
\includegraphics[width=3.2in,height=2.5in]{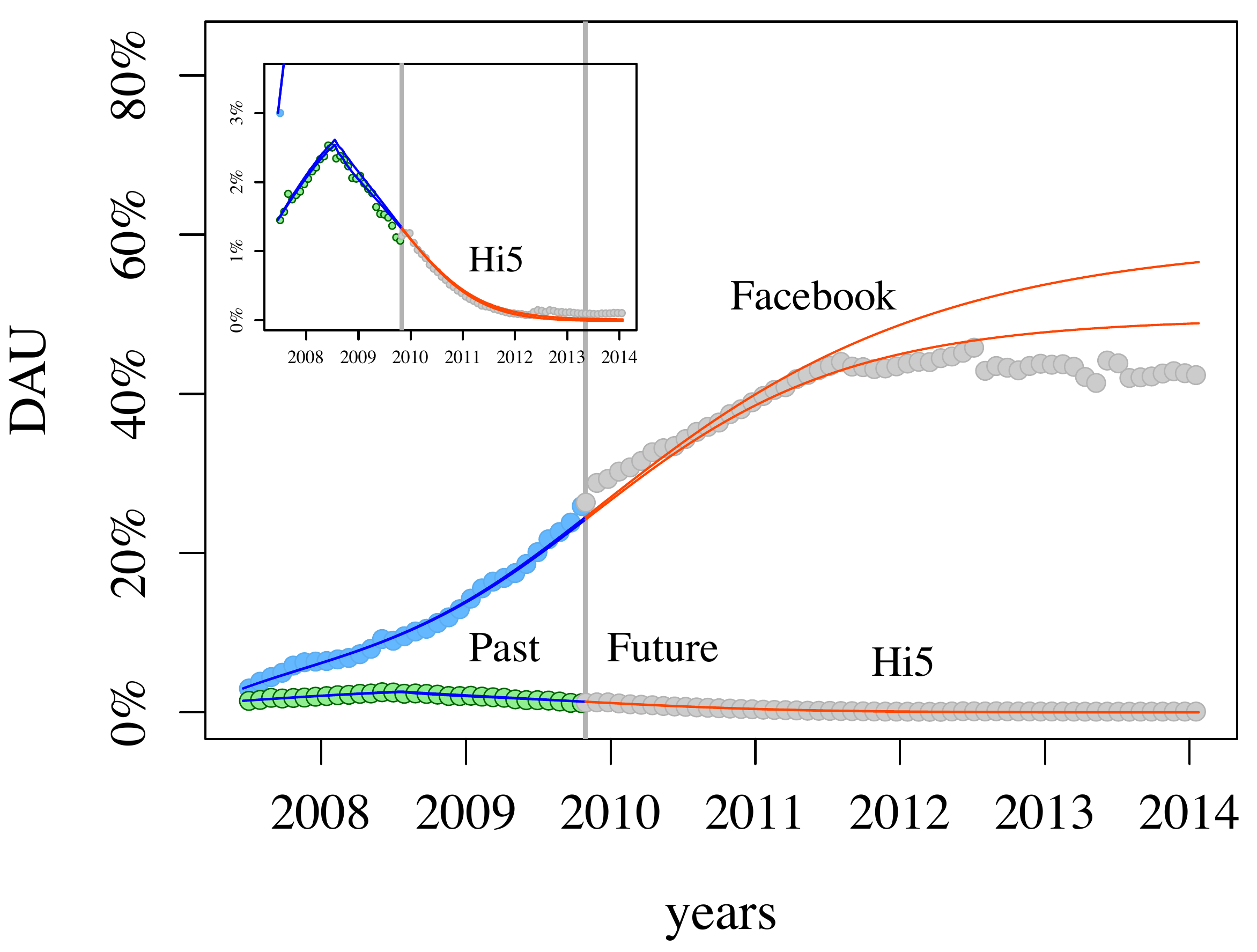}
\end{center}
\vspace{-8pt}
\caption{Model fit and predictions for the competition Facebook v.s.\ Hi5.\label{f:Hi5r}}
\end{figure}

\paragraph{Facebook v.s.\ Hi5}
Figure~\ref{f:Hi5r} shows the results of the model fit using the first two years (24 months) of the DAU time series in the competition between Facebook and Hi5.
The model fit shows great agreement with the data, blue line shows the model fit and red line shows its prediction.
The model was able to capture the sharp elbow near July 20, 2008.
Here the model estimates the total  (unaware + active + inactive) Facebook population at $C_{10}+C_{11} = 59\%$ and Hi5's population at $C_{01}+C_{11} = 5\%$; the model estimates that Hi5 user base grew largely due to media \& marketing campaigns and that Facebook's growth was through word-of-mouth;  the July 20, 2008 event largely reduced the average time spent on Hi5 by 95\%.
The model fit also estimates $\zeta = 3.7$, that is, by 2008 Hi5 users were almost four times as likely to join Facebook than the average Facebook adopter.

\begin{figure}[!!htb]
\begin{center}
\includegraphics[width=3.2in,height=2.5in]{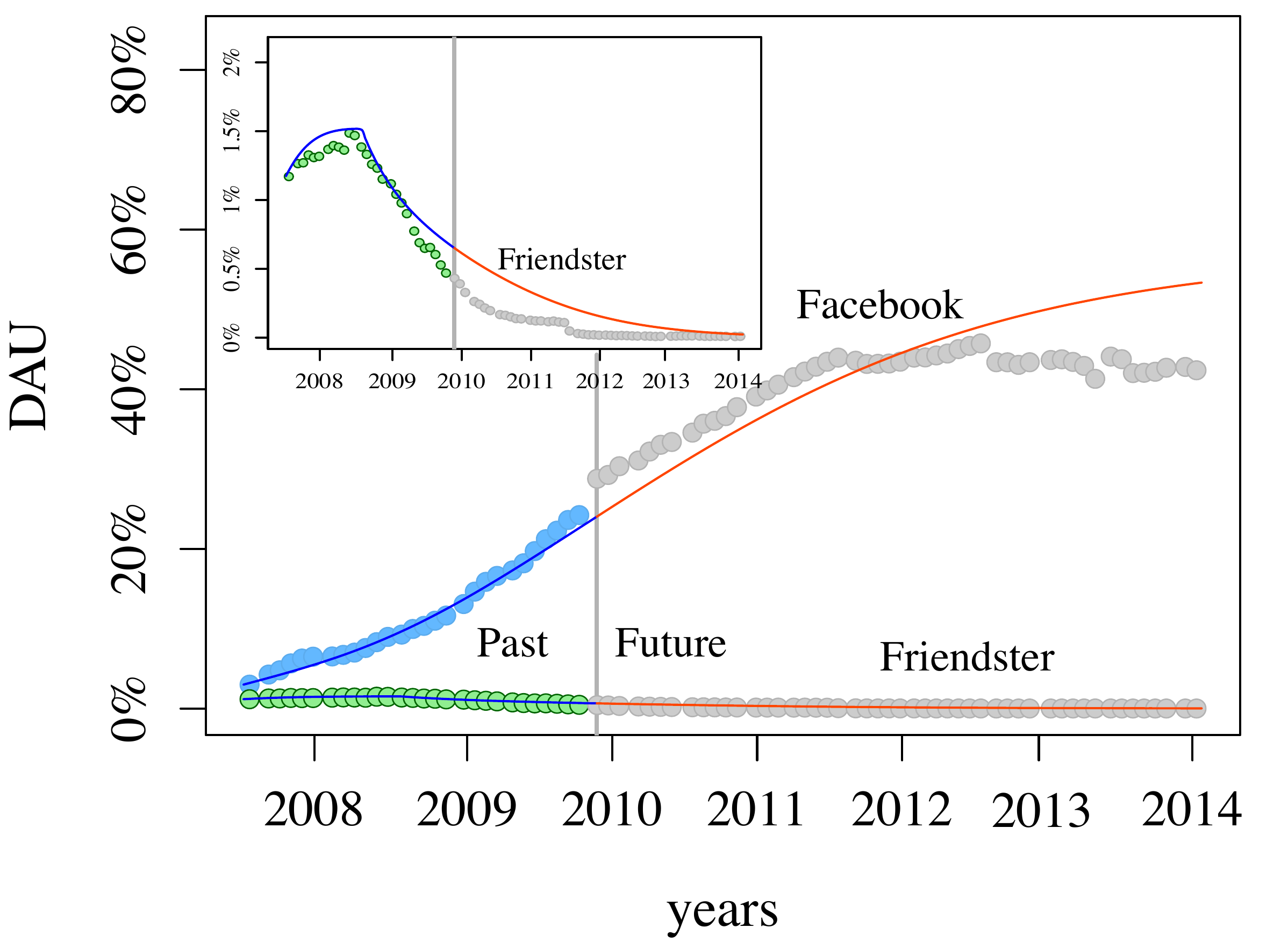}
\end{center}
\vspace{-8pt}
\caption{Model fit and predictions for the competition Facebook v.s.\ Friendster.\label{f:frier}}
\end{figure}

\pagebreak
\paragraph{Facebook v.s.\ Friendster}
Figure~\ref{f:frier} shows the results of the model fit using the first two years (24 months) of the DAU time series in the competition between Facebook and Multiply.
The model shows good agreement with the data both in the fitting phase (blue line) and the prediction phase (red line).
The model estimates the total  (unaware + active + inactive) Facebook population at $C_{10}+C_{11} = 59\%$ and Multiply's population at $C_{01}+C_{11} = 3.6\%$; the model estimates that both Facebook and Friendster user base grew largely due to word-of-mouth.
Immediately after the July 20, 2008 event the average time spent on Friendster is estimated to have decreased by nearly 99\%.
The model fit also estimates $\zeta = 1.8$ showing that Friendster users were almost twice as likely to join Facebook than the average Facebook adopter.

\section{Conclusions} \label{s:conclusions}

Our study sheds light onto the role of the attention-activity marketplace in the popularity (DAU time series) of membership-based websites.
Making use of the unique way by which Facebook affected its competitors in July, 2008, we derive a set of modeling principles that inform our proposed 
 attention-activity model design. 
Through a series of catalytic reactions that model user attention and activity interactions, together with media \& market diffusions, we propose a model that well captures the popularity competition between websites.
We fit the model parameters to real-world DAU time series data and show that our model not only fits well the DAU data but can also predict its future evolution.

In a 1969 lecture Herbert A.\ Simon warned us that information systems that help us generate more content than they help us reduce our time consuming such content would exacerbate the scarcity of attention~\cite{Simon}.
Our work takes a positive step towards modeling such phenomenon in Internet companies, providing insights into the connections between website popularity and user attention.
The model shows that two competing websites can co-exist without interfering with each other as long as users have enough attention to spare; this agrees with our data showing that before its new attention-demanding ``Wall'' feature Facebook did not seem to interfere with the popularity of MySpace, Hi5, Friendster, or Multiply.
Conversely, the model shows that websites fiercely compete when they share a sizable population of attention-starved users, and that such population can play a central role into the negative attention feedback loop that leads to the death of a website.
The model shows, for instance, that the popularity of a website with a large user base of tech savvy users -- or novelty-driven teen users -- can be easily preyed upon by a competing website, thus reducing the long-term viability of such websites.

Our hope is that further research in this direction will provide a better picture of the attention-activity marketplace, helping the design of information systems that do not need to overload the finite attention capacity of its users in order to survive.

\section{Acknowledgments}
This work was supported by NSF grant CNS-1065133 and ARL Cooperative Agreement W911NF-09-2-0053. The views and conclusions contained in this document are those of the author and should not be interpreted as representing the official policies, either expressed or implied of the NSF, ARL, or the U.S. Government. The U.S. Government is authorized to reproduce and distribute reprints for Government purposes notwithstanding any copyright notation hereon.

\balance

\end{document}